\documentclass[iop]{emulateapj}

\usepackage{graphicx}

\shorttitle{Origin of Cosmic Magnetic Fields}
\shortauthors{Cho}

\begin{document}

\title{Origin of Magnetic Field in the Intracluster Medium: Primordial or Astrophysical?}
\author{Jungyeon Cho\altaffilmark{1}}
\altaffiltext{1}{Department of Astronomy and Space Science,
       Chungnam National University, Daejeon, Korea; jcho@cnu.ac.kr} 


\begin{abstract}
The origin of magnetic fields in clusters of galaxies
is still an unsolved problem, which is largely due to our poor understanding of initial seed magnetic fields.
If the seed magnetic fields have primordial origins, it is likely that
large-scale pervasive magnetic fields were present before the formation of the large-scale
structure.
On the other hand, if they were ejected from astrophysical bodies, 
they were highly localized in space at the time of injection.
In this paper, using turbulence dynamo models for high magnetic Prandtl number fluids, we find constraints on the seed magnetic fields.
The hydrodynamic Reynolds number based on the Spitzer viscosity in the intracluster medium (ICM) is believed to be less than $O(10^2)$, while 
the magnetic Reynolds number can be much larger than that.
In this case, 
if the seed magnetic fields have primordial origins, they  
should be stronger than $O(10^{-11})$G, which is very close to
the upper limit of $O(10^{-9})$G set by the cosmic microwave background (CMB) observations.
On the other hand, if the seed magnetic fields were ejected from astrophysical bodies,
any seed magnetic fields stronger than $O(10^{-9})$G can safely magnetize the intracluster medium.
Therefore, it is less likely that primordial magnetic fields are the direct
origin of present-day magnetic fields in the ICM.

\end{abstract}
\maketitle

\section{Introduction}

Observations have it that magnetic fields of order $\mu$G are present 
in clusters of galaxies (see, for example, Kronberg 1994; Zweibel \& Heiles 1997; Carilli \& Taylor 2002; Widrow 2002; Govoni \& Feretti 2004; Vogt \& En{\ss}lin 2005; Ferrari et al. 2008;
Ryu et al. 2012).
No physical effect, e.g. a battery effect, seems to produce such strong magnetic fields directly.
Therefore, it is natural to assume that weak seed magnetic fields were produced 
at some time
and they have been
amplified later to current strengths.
According to this point of view, the origin of cosmic magnetism now involves
two separate issues:
generation of the initial seed magnetic fields and their amplification.
In this paper, we try to find constraints on the seed magnetic fields by studying
their amplification.

Theoretically, it is possible to generate weak magnetic fields in the early universe
before the recombination 
(see Widrow et al. 2012 and references therein).
However, it is highly unclear whether or not they have survived in the early universe and 
have become the seeds for present-day magnetic fields in the intracluster medium (ICM).
Probably, one of the best-known mechanisms for producing cosmic seed magnetic fields is the Biermann battery effect (Biermann 1950),
which should operate during the formation of the large-scale structure and can produce
a seed field of order $10^{-20}$G.
If seed magnetic fields were produced either in the early universe or during the large-scale structure formation,
it is likely that the seed fields were spatially homogeneous on the scale of galaxy clusters.

Magnetic fields expelled from astrophysical bodies
can also seed the ICM.  
First stars (Pudritz \& Silk 1989; Schleicher et al. 2010) or active galaxies (Hoyle 1969; Rees 1987; Daly \& Loeb 1990; Kronberg et al. 2001) can inject magnetized materials into the
intergalactic space.
Galactic winds, supernova explosions, ram pressure stripping can also provide magnetized materials to the intergalactic space (see, for example,  Rephaeli 1988; Kronberg 1994; Carilli \& Taylor 2002; Donnert et al. 2009;
 Arieli et al. 2011).
If a seed magnetic field was provided by an astrophysical object, it is likely that
the seed magnetic field was spatially localized at the time of injection.
As time goes on, the seed magnetic field can be dispersed  
away from the source and ultimately fill the whole system more or less homogeneously.
If the homogenization timescale is greater than the Hubble time, it is possible
to distinguish the primordial and the astrophysical origins of cosmic magnetism by observing intermittency
of magnetic field distribution.

Turbulence is commonly observed in clusters and filaments in simulations 
of the large-scale structure formation (Kulsrud et al. 1997;
Ryu et al. 2008; Xu et al. 2010; Vazza et al. 2011; Miniati 2014).
Turbulence can provide an efficient mechanism for amplification of weak seed magnetic fields.
In usual astrophysical environments, the magnetic diffusivity is very small, which enforces
fluid elements and magnetic field lines move together.
Therefore, chaotic turbulent motions can stretch magnetic field lines, 
making the magnetic field's strength and energy increase (Batchelor 1950; Zel'dovich et al. 1984; 
Childress \& Gilbert 1995).
This process is called small-scale turbulence dynamo.

Study of small-scale turbulence dynamo has a long history 
(e.g.~Batchelor 1950; Kazantsev 1968; Vainstein \& Ruzmaikin 1972;  
Pouquet et al. 1976; Meneguzzi et al. 1981; Kulsrud \& Anderson 1992; 
Cho \& Vishniac 2000; 
Haugen et al. 2004; Brandenburg \& Subramanian 2005; 
Schekochihin \& Cowley 2007; Cho et al 2009; Beresnyak 2012; Brandenburg et al. 2012; 
 Schober et al. 2012a; Bovino et al. 2013;
Yoo \& Cho 2014).
Most of the studies on the turbulence dynamo deal with
growth of a uniform or a homogeneous seed magnetic field in a fluid with unit magnetic Prandtl number $Pr_m$, which
is defined as $\nu/\eta$, where $\nu$ is the viscosity and $\eta$ is the 
magnetic diffusivity.
Nevertheless,
there are also studies for growth of localized seed magnetic fields (Molchanov et al. 1985; Ruzmaikin et al. 1989; Cho \& Yoo 2012; Cho 2013) in
unit magnetic Prandtl number turbulence and for growth of a uniform seed magnetic field in high magnetic Prandtl number turbulence (Schekochihin et al. 2004; Subramanian et al. 2006; Schekochihin \& Cowley 2007;  Bovino et al. 2013).

In this paper, we consider growth of both uniform and localized seed magnetic fields in  high
magnetic Prandtl number turbulence.
We will argue that, when we adopt turbulence dynamo models for 
high magnetic Prandtl number fluids,
astrophysical seed magnetic fields are more likely sources 
of magnetic fields in the ICM. 
In Section 2, we discuss turbulence dynamo in the ICM.
In Section 3, we describe numerical methods and, in Section 4, we present the results.
In Sections 5 and 6, we give discussions and summary, respectively.

\section{Turbulence dynamo in high magnetic Prandtl number fluids}
If the hydrodynamic Reynolds number, $Re$ ($\equiv Lv/\nu$, where $L$
is the outer scale of turbulence and $v$ is the r.m.s. velocity) is very large, turbulence dynamo is 
so efficient that any seed magnetic fields  follow virtually the same evolutionary track
after a very short initial exponential growth stage (see Appendix).
Therefore, in this case, it is difficult to 
find a constraint on the strength of the initial seed magnetic field.
However, it is believed that $Re$ in the ICM 
does not exceed $O(10^2)$. 
In this case,
turbulence dynamo may not be so efficient and the strengths of the seed field may matter.
In this section, we discuss turbulence dynamo in fluids 
with $Re$ $< O(10^2)$.
Note that, because the magnetic
  diffusivity is expected to be still very small in the ICM, the
  magnetic Prandtl number, $Pr_m$, can be much larger than unity (i.e.~$\nu \gg \eta$).

\subsection{The hydrodynamic Reynolds number, $Re$, in the ICM}
 Due to low density and high temperature,
the mean free path  between  collisions, $l_{mfp}$, is very large and, therefore, the viscosity, which is approximately $l_{mfp}$ times the particle velocity,
may not be negligibly small in the ICM. 
  Estimates for $Re$ in cool cores and hot ICM, do not exceed $O(10^2)$
  (Robinson 2004; Ruszkowski et al. 2004; Reynolds et al. 2005; Brandenburg \& Subramanian 2005; 
 Schekochihin \& Cowley 2006; Brunetti \& Lazarian 2007)
  if we use the Spitzer (1962)
formula for the viscosity.       

In fact, the hydrodynamic Reynolds number for the central $\sim$400 kpc region
 of a cluster can be
\begin{eqnarray}
 Re \approx 28 \left( \frac{v}{400 km/s} \right) 
            \left( \frac{L}{400~kpc} \right) 
            \left( \frac{ k_B T }{ 8~keV } \right)^{-2.5}   \nonumber \\
            \left( \frac{n}{ 0.001 cm^{-3} } \right)
            \left( \frac{\ln \Lambda}{40} \right)    \label{eq:reicm}
\end{eqnarray}
(modified from Equation (2) in Brunetti \& Lazarian 2007),
where $n$ is the number density,
 $k_B$ is the Boltzmann constant, $T$ is the temperature, and
$\ln \Lambda$ is the Coulomb logarithm.
In this paper,
we mainly assume that $Re \lesssim O(10^2)$ in the ICM.

\subsection{Theoretical expectations}
If $Re \lesssim 10^2$, the viscous-cutoff wavenumber $k_d$, at which 
the velocity spectrum drops quickly due to viscous damping, is not much larger than the
driving wavenumber.
Therefore, the eddy turnover time at the viscous-cutoff scale is not much different from
the large-scale eddy turnover time $L/v$, which
 makes turbulence dynamo inefficient.
 Because of inefficient turbulent dynamo, 
 the strength of the initial seed magnetic field becomes important.
In this section we consider turbulence dynamo in fluids with high magnetic Prandtl numbers\footnote{
   In this paper, by a high magnetic Prandtl number, we mean
   the hydrodynamic Reynolds number is less than $\sim O(10^2)$ and the magnetic Reynolds number is
   much larger than the hydrodynamic Reynolds number.}.

At the exponential growth stage, the magnetic field lines are stretched by 
turbulent motions at the viscous-cutoff scale,
also known as the Kolmogorov scale.
Since the viscous-cutoff scale is close to the outer scale of turbulence in high magnetic  Prandtl number fluids, 
the kinetic energy density at the viscous-cutoff scale is not much smaller than that of the outer scale.
Therefore, the exponential growth stage lasts for a long time until the magnetic energy density
becomes comparable to the kinetic energy density at the outer scale of turbulence.
In other words, the system is in exponential growth stage most of the time in high magnetic
Prandtl number turbulence\footnote{
  There might be a short linear growth stage after the exponential stage (see Cho \& Yoo 2012).
  But, existence of the linear growth stage is not important in our current discussions.}.

During the exponential growth stage, the magnetic energy density follows
\begin{equation}
   B^{2}(t) \propto B_0^2 \exp(t/\tau_d),
\end{equation}
where $B_0$ is the strength of the mean magnetic field and 
$\tau_d$ is proportional to the eddy turnover time at the viscous-cutoff scale:
$ \tau_d \propto l_d/v_d$. 
The duration of the exponential growth stage is
\begin{eqnarray}
    t \propto   \tau_d \ln(v_d^2 / B_0^2)  \\ \nonumber
       \propto  (L/v) Re^{-1/2} \ln(v^2 Re^{-1/2} /B_0^2 ), \\ \nonumber
\end{eqnarray}
where 
we use $v_l \propto l^{1/3}$ and $L/l_d \sim Re^{3/4}$.
For a fixed $Re$, we get
\begin{equation}
  t \propto C_1 - C_2 \ln{B_0},   \label{eq:t}
\end{equation}
where $C_1$ and $C_2$ are constants that depend on $Re$.
It is obvious from this equation that a weaker seed magnetic field takes more time
to reach saturation. 
Therefore, if a seed magnetic field is so weak that the exponential growth stage
takes more than the Hubble time, it cannot be the origin of cosmic magnetic fields.

If $Re \sim O(1)$, we get
\begin{equation}
   t \sim (L/v) \ln{ B_{eq}^2/B_0^2 },  \label{eq:t1}
\end{equation}
where $B_{eq}$  is the equipartition magnetic field (i.e.\  $B_{eq}^2/4 \pi = \rho v^2$),
 which is  of order $10^{-5}$G in typical clusters.
In this case, a seed magnetic field whose strength is weaker than
\begin{equation}
  B_{0, crit} \sim B_{eq} \exp( -t_{H}v/2L ),
\end{equation}
where $t_H$ is the age of the universe,
cannot explain present-day magnetic fields in the ICM.
In a cluster with $L/v \sim 10^{9}$ years and $B_{eq} \sim 10\mu$G, the critical strength would be
\begin{equation}
          B_{0, crit} \sim 0.1 \mu G.
\end{equation}
Note, however, that this estimate is highly uncertain because 
we ignored constants of order unity in many places while deriving Equation (\ref{eq:t1}).
In actual ICM, it is more complicated to obtain $B_{0, crit}$ because $Re$
can be greater than $O(1)$. Therefore, we need numerical simulations to get better estimates
for $B_{0, crit}$.

\section{Numerical methods}
\subsection{Numerical code}
We use a pseudospectral code to solve the 
incompressible magnetohydrodynamic (MHD) equations in a periodic box of size $2\pi$ ($\equiv L_{sys}$):
\begin{eqnarray}
\frac{\partial {\bf v} }{\partial t} =(\nabla \times {\bf B})
        \times {\bf B} -(\nabla \times {\bf v}) \times {\bf v}
      + \nu \nabla^{2} {\bf v} + {\bf f} + \nabla P^\prime ,
        \label{veq}  \hspace{5mm} \\ 
\frac{\partial {\bf B}}{\partial t}= 
     \nabla \times ({\bf v} \times{\bf B}) + \eta \nabla^{2} {\bf B} ,
     \label{beq}  \hspace{10mm}
\end{eqnarray}
where 
$ 
      \nabla \cdot {\bf v} =\nabla \cdot {\bf B}= 0,
$ 
$\bf{f}$ is random driving force,
$P'\equiv P + v^2/2$, ${\bf v}$ is the velocity,
and ${\bf B}$ is the magnetic field divided by $(4\pi \rho)^{1/2}$.
We use 22 forcing components in the wavenumber range $2 < k <\sqrt{12}$, which means the driving scale $L$ is
$\sim L_{sys}/2.5$.
In our simulations, 
$v\sim$1 before saturation.
Therefore, 
the large-scale eddy turnover time,  
$\sim L/v$, is approximately $\sim$2.5 before saturation.  
Other variables have their usual meaning.
We use $512^3$ grid points.

We use normal viscosity for the velocity dissipation term, $\nu \nabla^2 {\bf v}$, 
and hyperdiffusion for the magnetic dissipation term. 
The power of hyperdiffusion
is set to 3, such that the magnetic dissipation term in Equation (\ref{beq}) 
is replaced with
$ 
 \eta_3 (\nabla^2)^3 {\bf B}.
$
Therefore, the magnetic Prandtl number ($=\nu/\eta$) is larger than one.
Since we use hyperdiffusion, 
dissipation of magnetic field is negligible for small wavenumbers and 
it abruptly increases near $(2/3)k_{max}$, where $k_{max}$=256.

\subsection{Initial conditions}

At $t=0$, turbulence is fully developed and
either a uniform or a localized seed magnetic field gets ``switched on''.
For localized seed magnetic fields, we consider two shapes: tube-like 
and doughnut-like seed magnetic fields.

For tube-shaped seed magnetic fields, 
we use the following expression for the magnetic field at t=0:
\begin{equation}
  {\bf B}(r_{\bot}) 
  = B_{max}  e^{ -r_{\bot}^2 /2 \sigma_0^2 } 
     \hat{\bf x},   \label{eq:b_tube}
\end{equation}
where $\sigma_0=8$, $r_{\bot}=( \Delta y^2+ \Delta z ^2 )^{1/2}$, 
and $\Delta y$ and $\Delta z$ are the distances measured from the
tube in grid units. 
The unit vector $ \hat{\bf x}$ is parallel to the x-axis.
The maximum strength of the magnetic field at t=0 is $B_{max}$.

For doughnut-shaped seed magnetic fields, 
we use the following expression for the magnetic field at t=0:
\begin{equation}
  {\bf B}(\Delta x, r_{\bot}) 
  = \frac{ B_{max} }{ 2 \sigma_0^2 e^{-1}}  r_{\bot} ^2 e^{ -r_{\bot}^2 /2 \sigma_0^2 } 
      e^{ -\Delta x^2/8 \sigma_0^2 } 
     \hat{\bf \theta}_{\perp},   \label{eq:b_doughnut}
\end{equation}
where $B_{max}=0.01$, $\sigma_0=4\sqrt{2}$, $r_{\bot}=( \Delta y^2+ \Delta z ^2 )^{1/2}$, 
and $\Delta x, \Delta y,$ and $\Delta z$ are distances measured from the center of the
numerical box in grid units. The unit vector $ \hat{\theta}_{\perp}$ is perpendicular to
$(\Delta x,0,0)$ and $(0,\Delta y, \Delta z)$.
The maximum strength of the magnetic field at t=0 is $B_{max}$.
Since $\sigma_0=4\sqrt{2}$ in Equation (\ref{eq:b_doughnut}), the size of
the magnetized region at t=0 is $\sim$16 in grid units, which is
$\sim$1/32 of the simulation box size. 
Therefore, in a cluster of size $\sim$1 Mpc, the size of the initially magnetized region corresponds 
to  $\sim$30 kpc.

\section{Results}

\subsection{Results for uniform seed magnetic fields}

\begin{figure*}
\center
\includegraphics[angle=0,width=0.48\textwidth, bb=100 140 500 550]{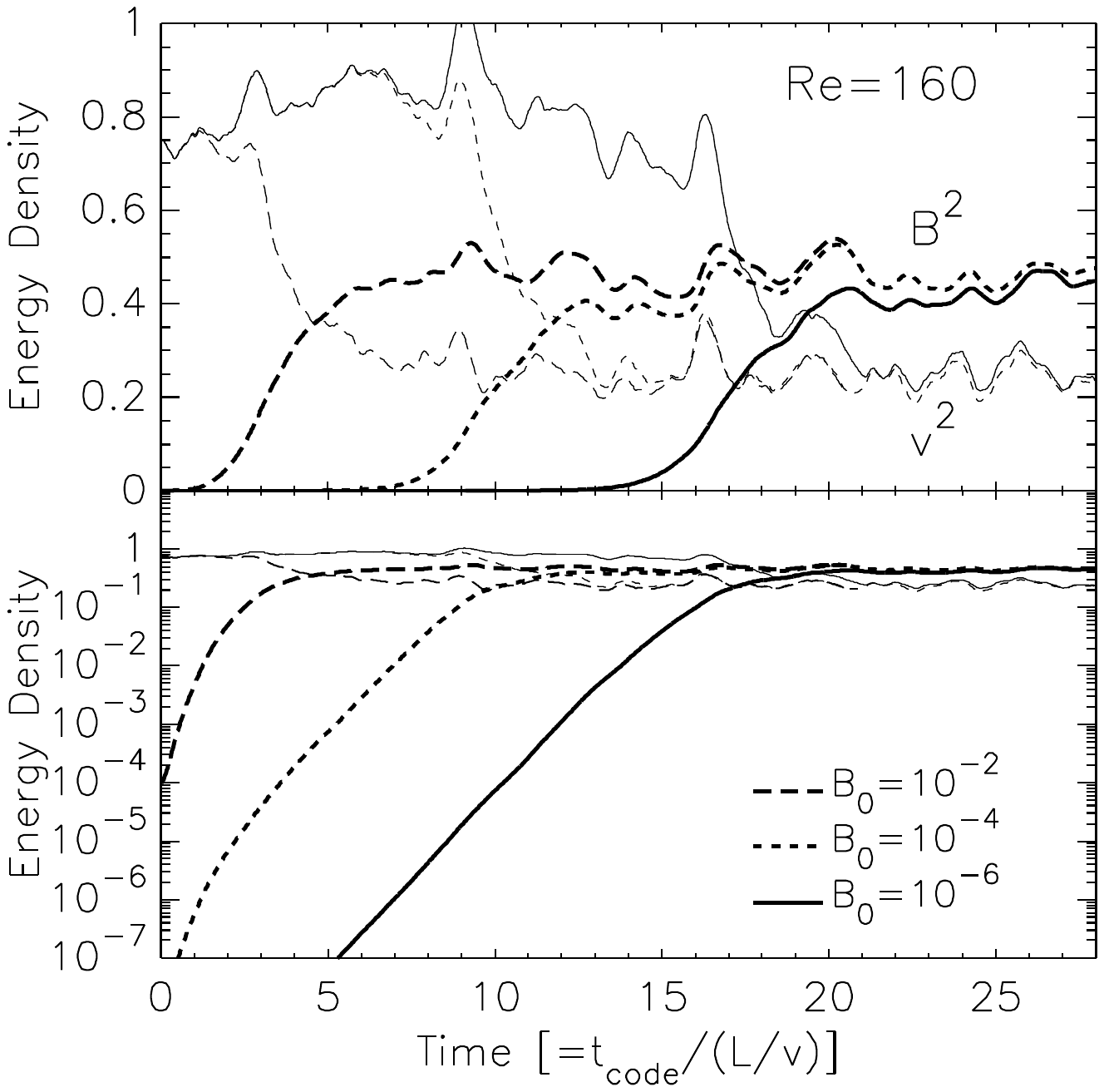}  
\includegraphics[angle=0,width=0.48\textwidth, bb=100 140 500 550]{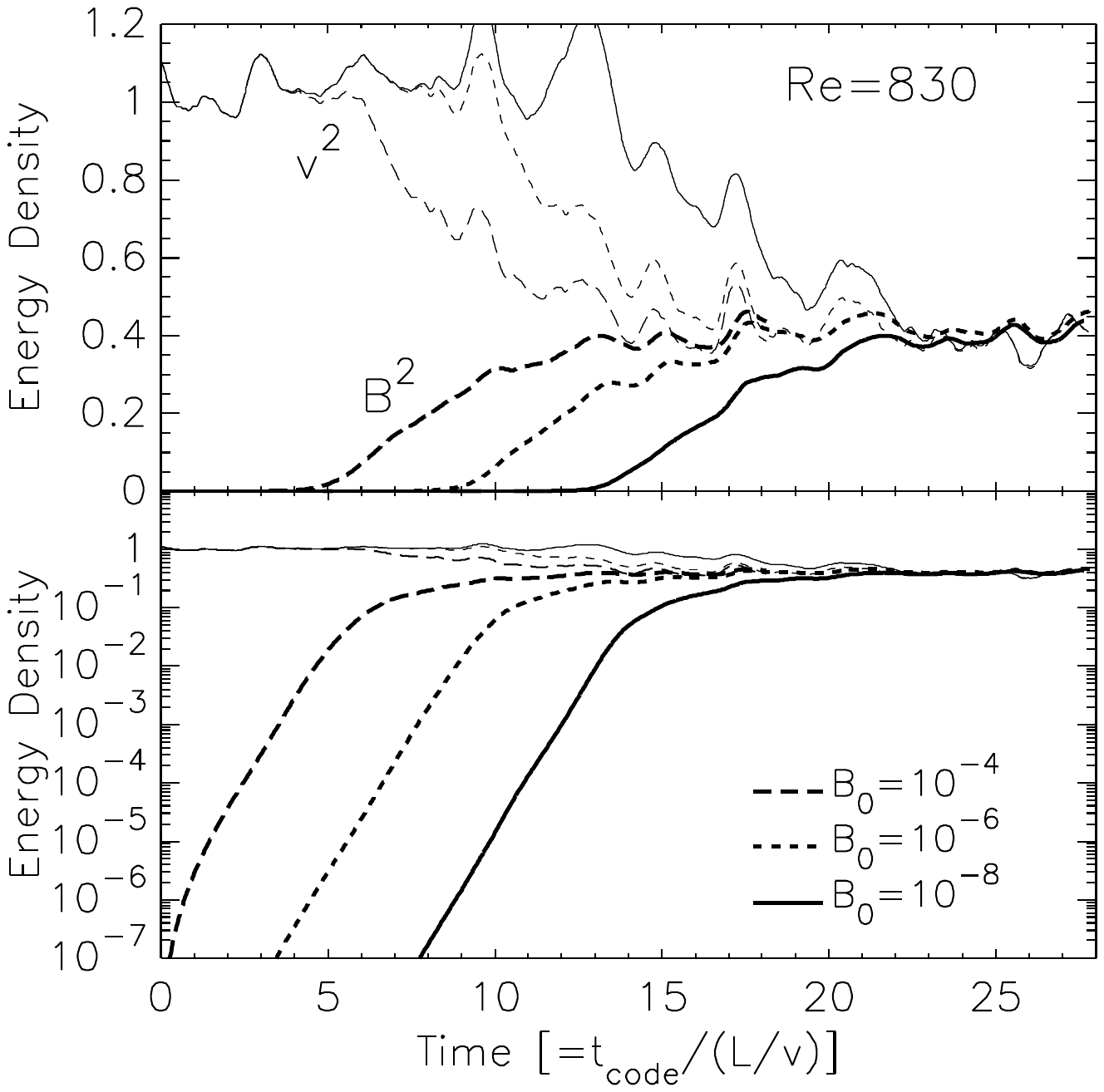} 
\caption{Growth of uniform seed magnetic fields in a high magnetic Prandtl number turbulence
with the hydrodynamic Reynolds number ($Re$) of $\sim$ 160. The scales for the vertical axes are different: linear in the upper
panel and logarithmic in the lower panel. 
The exponential growth stage takes up most of the 
growth time.
The growth time of $B^2$ depends on the strength
of the uniform seed field, $B_0$: a weaker seed field reaches the saturation stage later.
From Runs 512-R1-UB$_010^{-2}$, 512-R1-UB$_010^{-4}$, and 512-R1-UB$_010^{-6}$.
  }  \label{fig:unif1}
\caption{Growth of uniform seed magnetic fields in a high magnetic Prandtl number turbulence
with $Re \sim$ 830.  
It seems that the exponential growth stage is followed by a short linear growth stage.
Although the growth time of $B^2$ depends on the strength of the uniform seed magnetic field, $B_0$, 
the dependence is less pronounced compared with the results in Figure~\ref{fig:unif1}.
From Runs 512-R2-UB$_010^{-4}$, 512-R2-UB$_010^{-6}$, and 512-R2-UB$_010^{-8}$.
} \label{fig:unif2}
\end{figure*}

Figure~\ref{fig:unif1} shows the time evolution of $B^2$ and $v^2$.
In the simulations (Runs 512-R1-U's in Table 1), the viscosity is 0.015 and the driving scale is $L=L_{sys}/k_f \sim 2.5$ 
and the velocity before saturation is $\sim \sqrt{0.9}$.
Therefore, the hydrodynamic Reynolds number $Re$ based on the velocity before saturation is roughly 160.
Both the upper and the lower panels of Figure~\ref{fig:unif1} show the time evolution of the same quantities, 
but the scales for the vertical axes are different;
linear in the former and logarithmic in the latter.
Initially only the uniform component of the magnetic field exists and, as times goes on, turbulent motions
stretch magnetic field lines and make the magnetic field grow.
As the magnetic energy density increases, the kinetic energy density of turbulence decreases.
During the saturation stage, magnetic energy density is a bit higher than the kinetic energy density in all runs.

In Figure~\ref{fig:unif1} we can clearly see that the exponential growth stage takes up most of the growth time
(see the lower panel).
It is possible that the exponential growth stage is followed by a linear growth stage (see discussions in 
Cho \& Yoo 2012). However, because our $Re$ is too small and, therefore, we do not fully resolve
the inertial range, the duration of the linear growth stage, if any, should be very short.

In Figure~\ref{fig:unif1}, we can see that a weaker seed magnetic field takes more time to reach the saturation stage.
When the maximum strength of the seed magnetic field is $10^{-2}$, which means that the 
maximum strength of the initial magnetic field is roughly $10^{-2}$ times the equipartition magnetic
 field strength,
the system reaches saturation within $t\sim 5(L/v)$.
However, if the maximum strength of the seed magnetic field is $10^{-6}$, 
the system reaches saturation within $t\sim 20(L/v)$.

Consider a cluster with $L\sim$400 kpc, $v\sim$ 400 km/s and equipartition magnetic field of 
$\sim 10\mu$G.
In this case, the large-scale eddy turnover time is 
\begin{equation}
   L/v  \sim 10^{9} \mbox{~years}.  \label{eq:eddytt}
\end{equation}
If the strength of the initial seed magnetic field is $0.01$ nG, which is $10^{-6}$ times the equipartition magnetic field strength, the system reaches saturation $\sim$20 billion years after the big bang.
This simple argument implies that any primordial seed magnetic field weaker than $\sim 0.01$nG
cannot be the direct origin of cosmic magnetism.
Note that
the large-scale eddy turnover time in Equation~(\ref{eq:eddytt}) 
is in rough agreement with the values in
some numerical simulations of the large-scale structure formation (see, for example, Ryu et al. 2008).

 As aforementioned, we expect that $Re \lesssim 10^2$ in the ICM (see Equation (\ref{eq:reicm})).
Nevertheless, it could be possible that $Re > 10^2$ in some clusters.
Therefore, it is worth investigating
turbulence dynamo for a higher $Re$.  
In Figure~\ref{fig:unif2} we present results of simulations for a higher $Re$.
In the simulations (Runs 512-R2-U's in Table 1), the viscosity is 0.003,
the driving scale is $L=L_{sys}/k_f \sim 2.5$,
and the velocity before saturation is $\sim 1.0$, so that
$Re$ based on the velocity before saturation is roughly 830.
As in Figure \ref{fig:unif1}, we can clearly see the exponential growth stage and,
now, we can see that a short, but clearly visible, linear growth stage follows the exponential growth
stage.
As in Figure~\ref{fig:unif1}, a weaker seed magnetic field takes more time to reach saturation.
But now the growth rate of magnetic field is higher because the eddy turnover time at the
viscous-cutoff scale is shorter.
Therefore, compared with runs in Figure~\ref{fig:unif1}, somewhat weaker seed magnetic fields can be the direct origin
of magnetic field in the ICM.
Figure~\ref{fig:unif2} shows that a seed magnetic field as weak as $10^{-8}$ times the
equipartition strength can reach saturation
within $\sim 20(L/v)$.
If we consider the same cluster as before, 
a seed magnetic field as weak as $10^{-4}$nG can reach saturation
$\sim 20$ billion years after the big bang.

\subsection{Results for localized seed magnetic fields}

\begin{figure}[!ht]
    \begin{minipage}{0.95\columnwidth}
      \includegraphics[width=0.99\columnwidth]{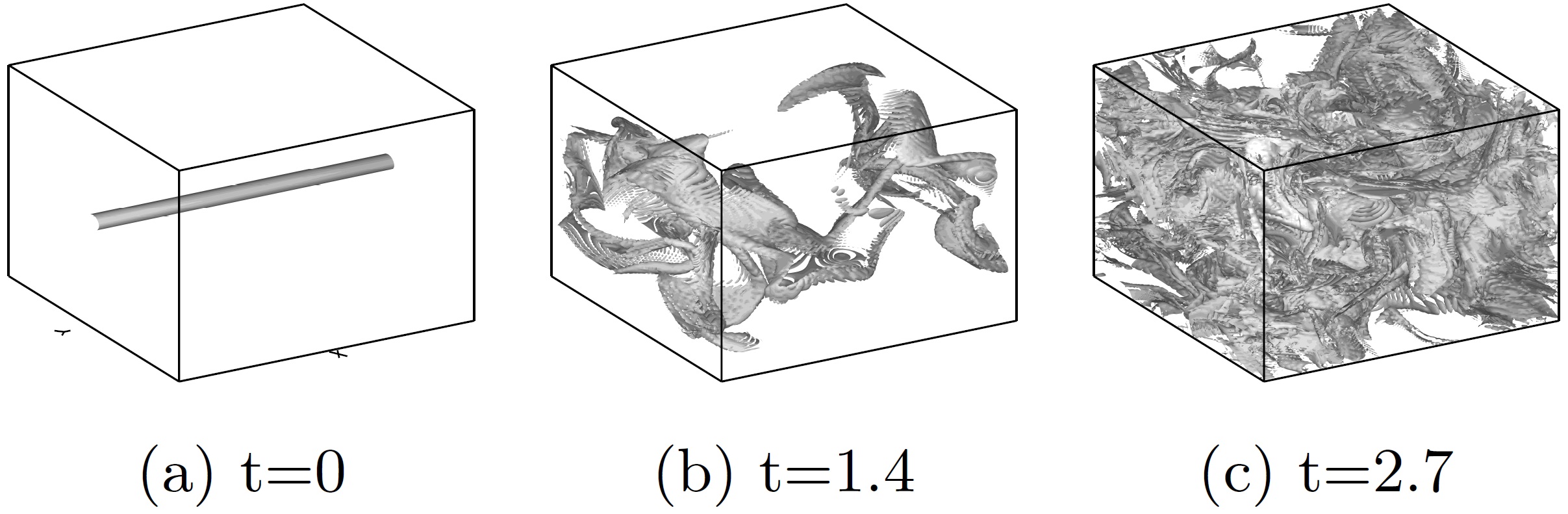}
      \caption{Homogenization of a tube-shaped localized seed magnetic field
      in high magnetic Prandtl number turbulence.
      The expansion of the magnetized region happens very fast.
      After homogenization, the subsequent evolution should be very similar to
      that of a uniform seed magnetic field case.
      The driving scale is about 2.5 times smaller than the size of  the computational box.
      In the shaded regions, the magnetic field is stronger than 0.1 times $B_{max}$.
      From Run 512-R1-TB$_{max}10^{-2}$.
      }\label{fig:tubejpg}
    \end{minipage}
    \hfill
    \begin{minipage}{0.95\columnwidth}
      \includegraphics[width=0.95\columnwidth, bb=165 160 430 570]{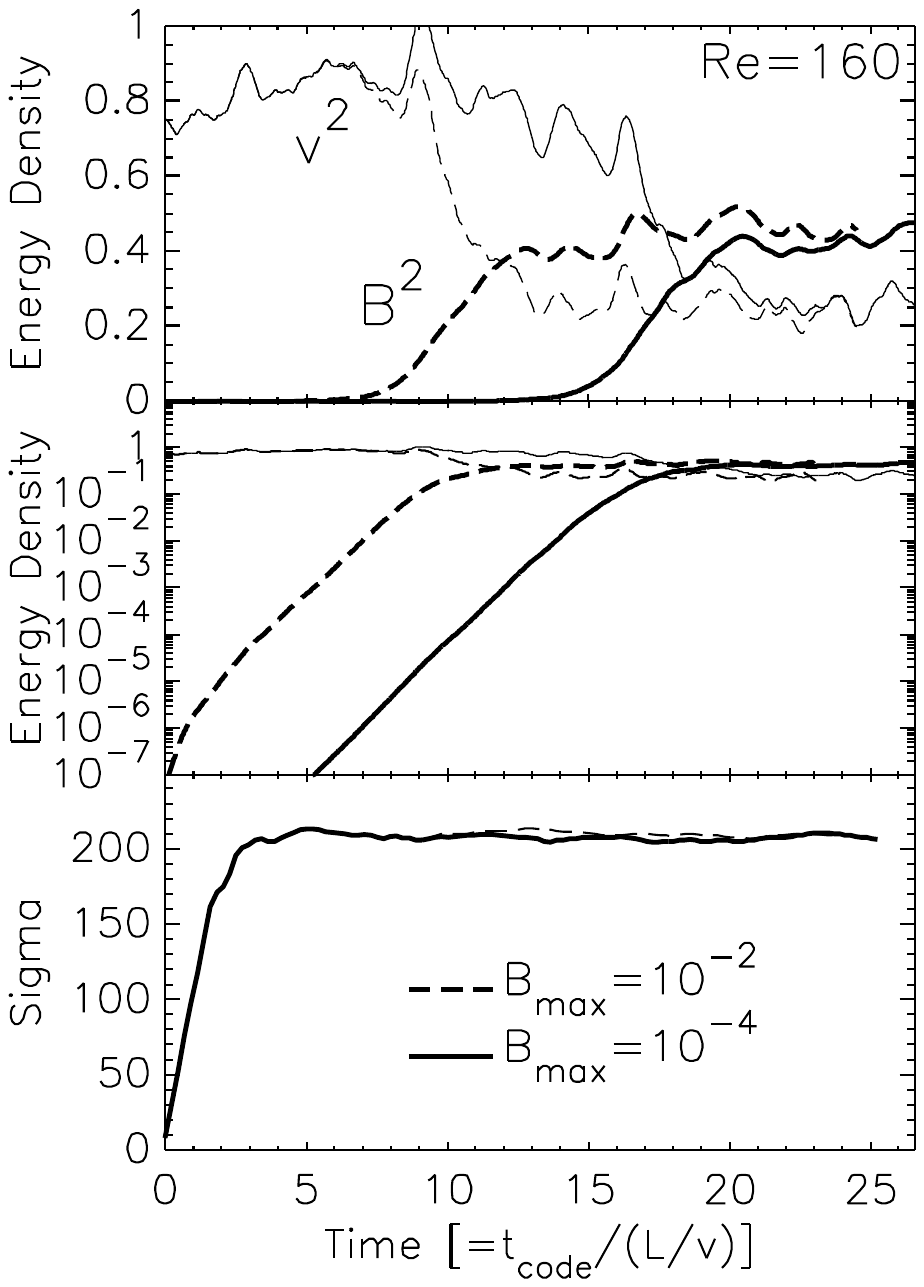}
      \caption{Growth of tube-shaped localized seed magnetic fields in high magnetic Prandtl number turbulence.
      The upper and the middle panels show the time evolution of the same quantities, $B^2$ and $v^2$, but
      their vertical scales are different: linear in the upper
      panel and logarithmic in the middle panel. 
       The lower panel shows the time evolution of the standard deviation
      of the magnetic field distribution, $\sigma$.
      In the lower panel, the two curves for $B_{max}=10^{-2}$ and $10^{-4}$
      almost coincide.
      From Runs 512-R1-TB$_{max}10^{-2}$ and 512-R1-TB$_{max}10^{-4}$.
         }\label{fig:tubet}
    \end{minipage}
\end{figure}

In this subsection, we investigate how localized seed magnetic fields grow in
 high magnetic Prandtl number turbulence.
 In all simulations in this subsection 
 (Runs 512-R1-T's and 512-R1-D's in Table 1),
the viscosity is 0.015 and the driving scale is $L=L_{sys}/k_f \sim 2.5$ 
and the velocity before saturation is $\sim \sqrt{0.9}$, so that
$Re$ based on the velocity before saturation is roughly 160.
If the seed magnetic field is localized in space, 
turbulent motions  stretch and disperse magnetic field  lines, which 
makes the magnetic field grow and, at the same time, the magnetized region expand
 in high magnetic Prandtl number turbulence 
as in unit magnetic Prandtl number turbulence.

Figure~\ref{fig:tubejpg} shows that the homogenization of a tube-shaped seed magnetic field is very fast.
The figure shows that the whole numerical box becomes magnetized within $\sim 2.7 (L/v)$ 
(see more quantitative results
in the lower panel of Figure~\ref{fig:tubet}).
In our simulations, the size of the computational box is $\sim 2.5L$.
Therefore, the observed rate of homogenization implies that
the speed at which the magnetized region expands is of order $v$ in high magnetic Prandtl number 
turbulence as in unit magnetic Prandtl number 
turbulence.

Figure~\ref{fig:tubet} shows the time evolution of $B^2$, $v^2$, and the standard deviation of the magnetic field distribution, which can be regarded as 
an approximate size of the magnetized region.
Both the upper and the middle panels of Figure~\ref{fig:tubet} show the time evolution of $B^2$ and $v^2$, 
but the scales for the vertical axes are different. 
Initially only the localized seed magnetic field exists and, as time goes on, turbulent motions
make the magnetic field grow.
As the magnetic energy density increases, the kinetic energy density of turbulence decreases.
The values of $B^2$ and $v^2$ at the saturation stage are very similar to those for uniform seed field cases
(compare Figures~\ref{fig:tubet} and \ref{fig:unif1}).

In the lower panel of Figure~\ref{fig:tubet}, we plot the time evolution of 
the standard deviation, $\sigma$, of magnetic field distribution:
\begin{eqnarray}
     \sigma = ( \sigma_y^2 + \sigma_z^2 )^{1/2}, \\
\    \sigma_i^2 =\frac{ \int ( x_i -\bar{x}_i)^2 |{\bf B}({\bf x},t)|^2 d^3x }{ \int  |{\bf B}({\bf x},t)|^2 d^3x },  \label{eq:sigma_i} \\
  \bar{x}_i=\frac{    \int  x_i  |{\bf B}({\bf x},t)|^2 d^3x }{ \int  |{\bf B}({\bf x},t)|^2 d^3x },   \label{eq:bar_i}
\end{eqnarray}
where $i$=y and z. 
Initially the standard deviation rises very quickly and the growth rate decreases after $\sim 1.5 (L/v)$.
Within $\sim 3(L/v)$, the standard deviation almost reaches the value for homogeneous distribution.
The behavior of $\sigma$ is not sensitive to the value of B$_{max}$: in fact, the curves for $B_{max}10^{-2}$ and
$B_{max}10^{-4}$ virtually coincide.

Since the seed magnetic field in Run 512-R1-T$B_{max}10^{-2}$ (dashed curve) is stronger than that in
Run 512-R1-T$B_{max}10^{-4}$ (solid curve), Run 512-R1-T$B_{max}10^{-2}$ reaches saturation earlier.
Note that Run  512-R1-T$B_{max}10^{-4}$ reaches saturation within $\sim 20 (L/v)$.
Therefore, for a cluster of galaxies with $L/v \sim 10^9$ years,
any tube-like seed magnetic field weaker than $\sim 10^{-4}$ times the equipartition strength
cannot produce a strong enough magnetic field within the Hubble time.
For example, 
if we consider the same cluster as before
($L_{sys}=$1 Mpc, $L=$400 kpc, 
 $v \sim 400$ km/s, $Re=160$, and the equipartition magnetic field strength of $\sim 10\mu$G), 
 it is likely that any localized tube-shaped seed magnetic field that is
 weaker than $\sim$1 nG cannot be the origin of magnetic field in the ICM.
 Of course, if $Re$ is larger, the minimum magnetic field strength will decrease.
 
\begin{figure}[!ht]
    \begin{minipage}{0.95\columnwidth}
      \includegraphics[width=0.99\columnwidth]{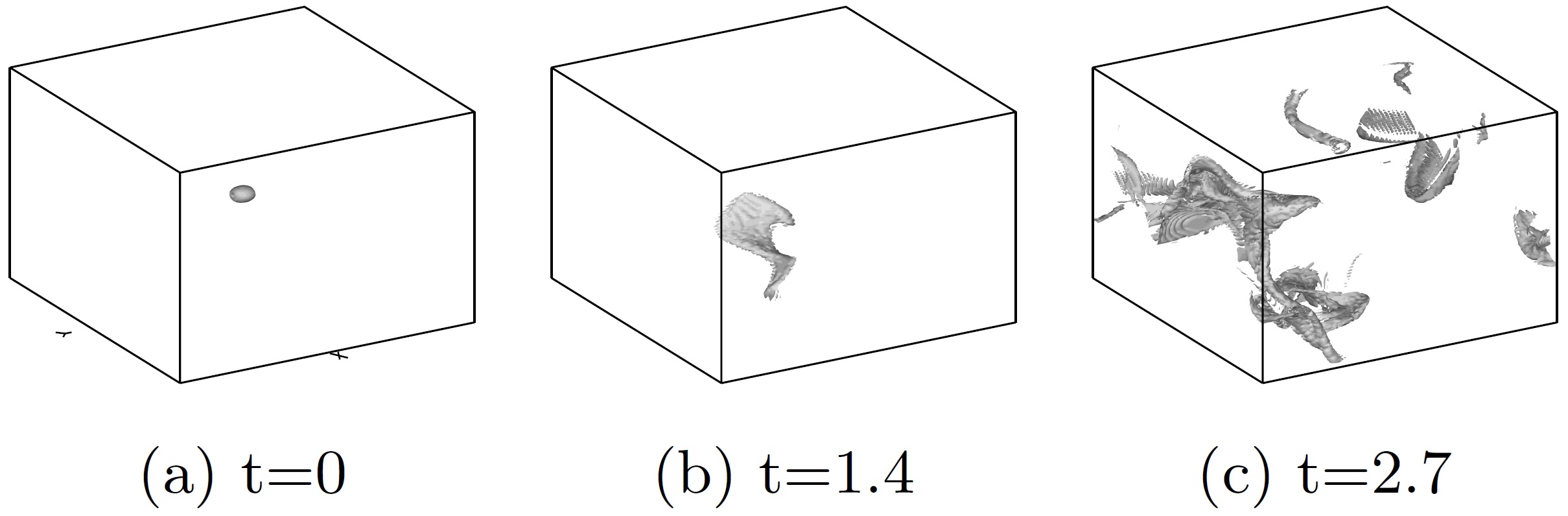}
      \caption{Homogenization of a doughnut-shaped localized seed magnetic field
      in high magnetic Prandtl number turbulence.
      The expansion of the magnetized region happens very fast.
      After homogenization, the subsequent evolution should be very similar to
      that of a uniform seed magnetic field case.
      The driving scale is about 2.5 times smaller than the size of  the computational box.
           In the shaded regions, the magnetic field is stronger than 0.1 times $B_{max}$.
            From Run 512-R1-DB$_{max}10^{-2}$.
      }\label{fig:bsprjpg}
    \end{minipage}
    \begin{minipage}{0.95\columnwidth}
      \includegraphics[width=0.95\columnwidth, bb=165 160 430 570]{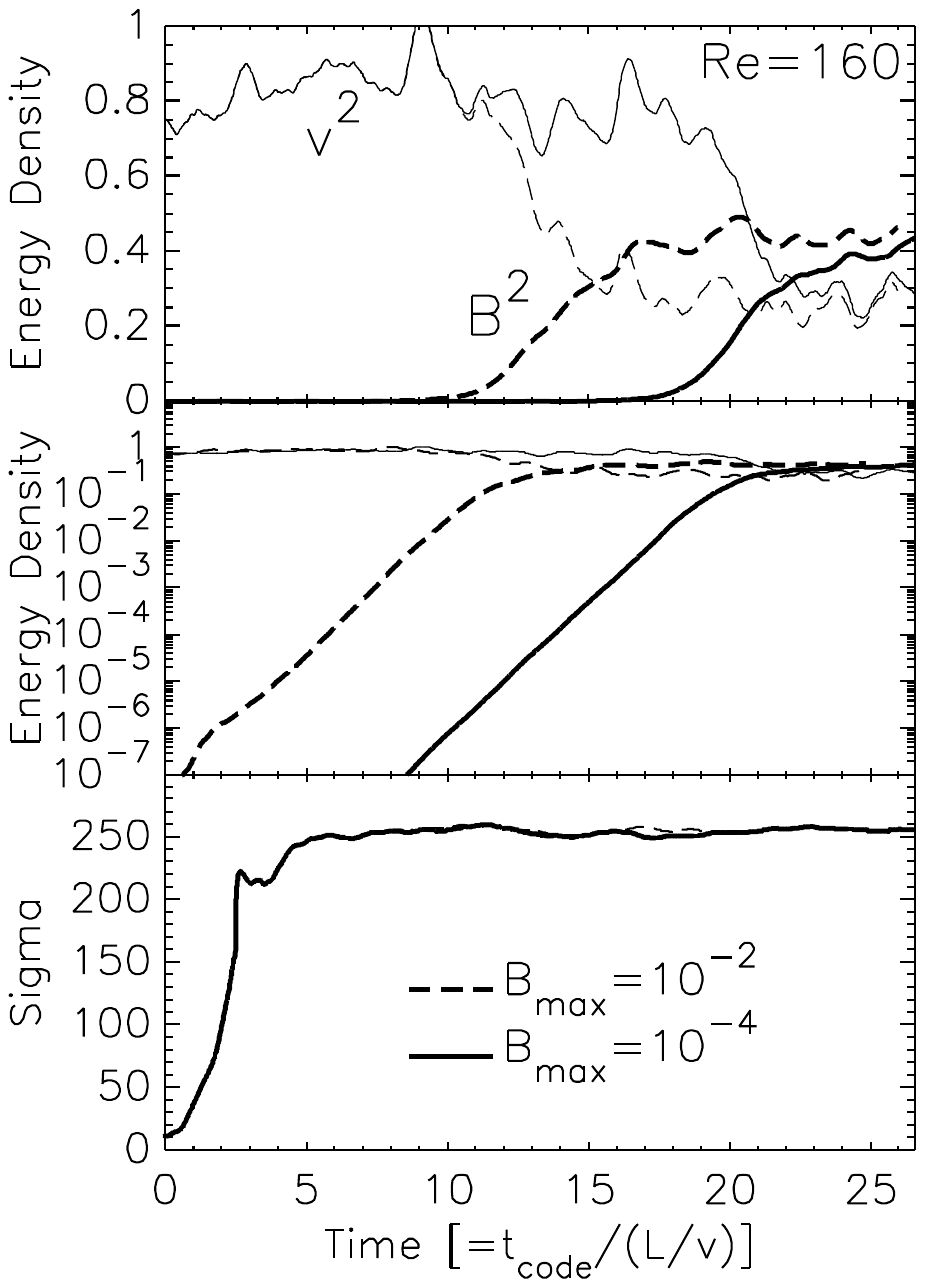}
      \caption{Growth of doughnut-shaped localized seed magnetic fields in high magnetic Prandtl number turbulence.
      The upper and the middle panels show the time evolution of the same quantities, $B^2$ and $v^2$. 
       The lower panel shows the time evolution of the standard deviation
      of the magnetic field distribution $\sigma$.
      In the lower panel, the two curves for $B_{max}=10^{-2}$ and $10^{-4}$
      almost coincide.
            From Runs 512-R1-DB$_{max}10^{-2}$ and 512-R1-DB$_{max}10^{-4}$.
      }\label{fig:bsprt}
    \end{minipage}
\end{figure}


Figure~\ref{fig:bsprjpg} shows that the homogenization time for a doughnut-shaped seed magnetic field is also very fast.
The figure shows that the magnetic field spreads out and fills the whole numerical box 
within $\sim 2.7 (L/v)$.
Actually, if we plot the time evolution of 
the standard deviation, $\sigma$, of magnetic field distribution,
\begin{equation}
     \sigma = ( \sigma_x^2+\sigma_y^2 + \sigma_z^2 )^{1/2},  \label{eq:sig3d}  
\end{equation}
where $\sigma_i^2$, $i=$x, y, and z, is defined as in Equation~(\ref{eq:sigma_i}), 
it reaches the value for homogeneous distribution after $\sim 4(L/v)$
(the lower panel of Figure~\ref{fig:bsprt}).
As in the tube-like seed magnetic field cases, 
the behavior of $\sigma$ is not sensitive to the value of B$_{max}$: the curves for $B_{max}10^{-2}$ and
$B_{max}10^{-4}$ almost coincide.
As we discussed above, the observed rate of homogenization is consistent with the fact that
the speed at which the magnetized region expands is of order $v$.
If we compare homogenization of a tube-like seed magnetic field (Figure~\ref{fig:tubejpg}) and
a doughnut-like seed magnetic field  (Figure~\ref{fig:bsprjpg}), the latter produces more intermittent 
magnetic field distribution during and at the end of the homogenization process.

Figure~\ref{fig:bsprt} shows the time evolution of $B^2$, $v^2$, and $\sigma$.
The Run  512-R1-D$B_{max}10^{-2}$ reaches saturation after $\sim 15 (L/v)$ and
the Run  512-R1-D$B_{max}10^{-4}$ reaches saturation after $\sim 25 (L/v)$.
Therefore, for a cluster of galaxies with $L/v \sim 10^9$ years,
any doughnut-like seed magnetic field weaker than $\sim 10^{-3}$ times the equipartition strength
cannot produce a strong enough magnetic field within the Hubble time.
For example, if we consider the cluster of galaxies mentioned earlier ($L_{sys}=$1 Mpc,  $L=$400 kpc,
 $v \sim 400$ km/s, $Re=160$, and equipartition magnetic field strength of $\sim 10\mu$G), 
 it is likely that any localized doughnut-shaped seed magnetic field whose maximum strength is
 weaker than $\sim$10nG cannot be the origin of magnetic field in the ICM.

\begin{figure}[!ht]
    \begin{minipage}{0.95\columnwidth}
      \includegraphics[width=0.99\columnwidth]{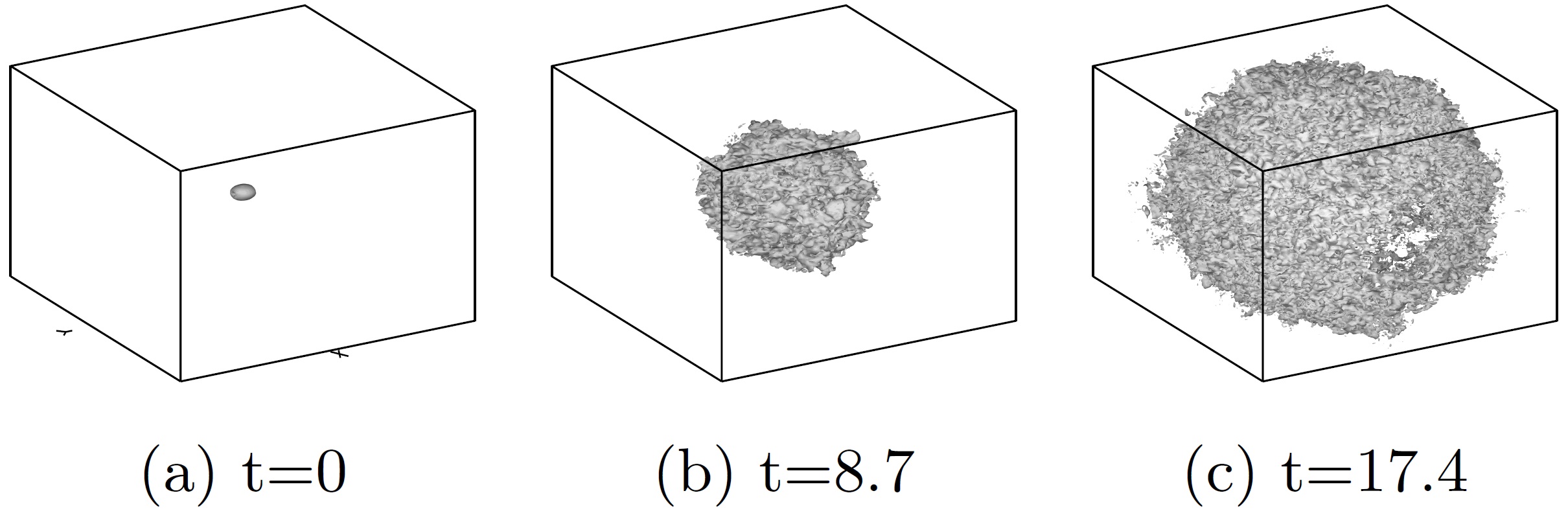}
      \caption{Same as Figure~\ref{fig:bsprjpg}, but 
      the driving scale is smaller: it is about 20 times smaller than a side of  the computational box.
            From Run 512-REF.
      }\label{fig:k20jpg}
    \end{minipage}
    \begin{minipage}{0.95\columnwidth}
      \includegraphics[width=0.95\columnwidth, bb=165 160 430 570]{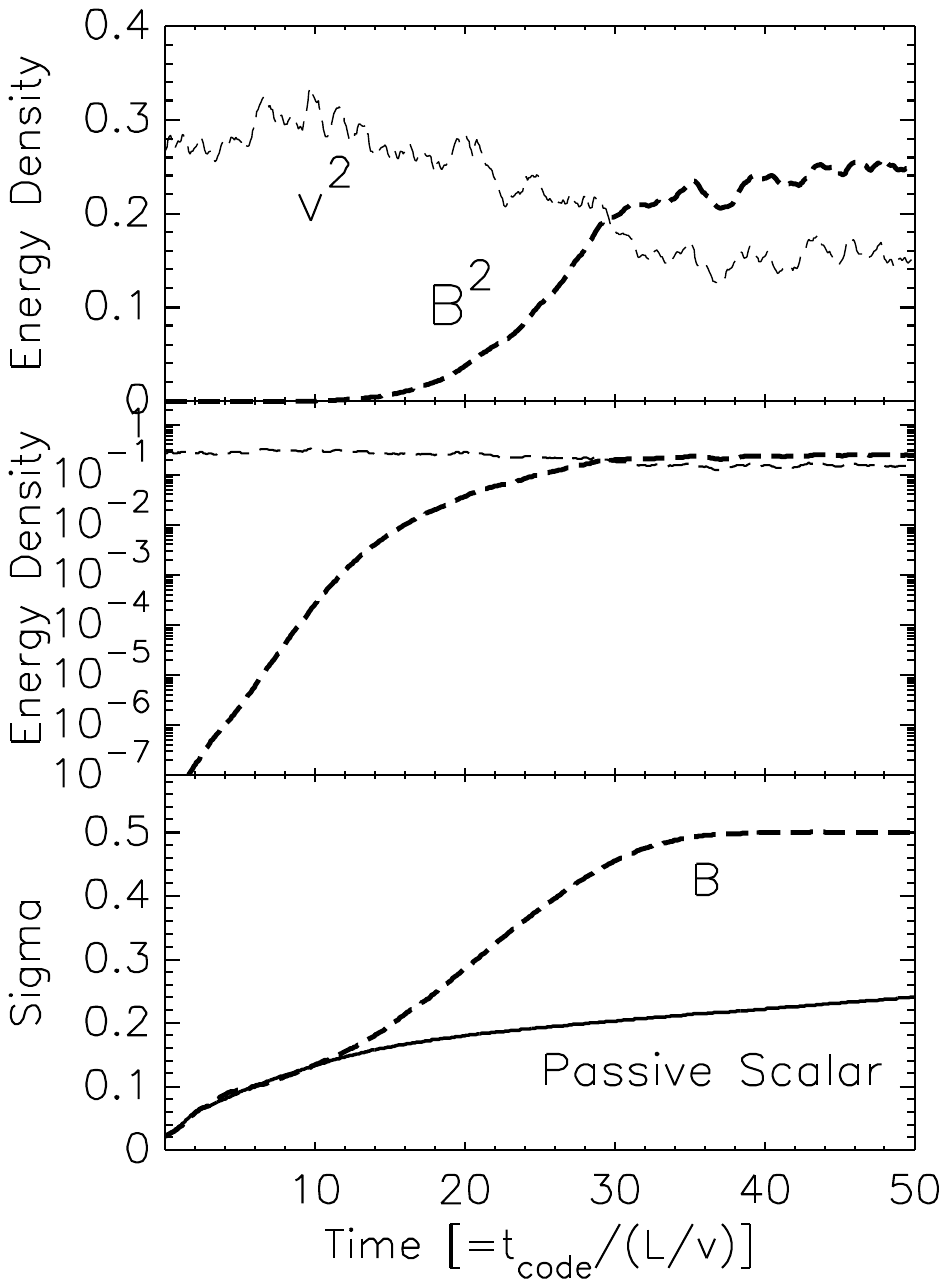}
      \caption{Growth of doughnut-shaped localized seed magnetic fields in high magnetic Prandtl number turbulence
      with a small-scale driving: the driving scale is about 20 times smaller than a side of  the computational box.
      The upper and the middle panels show the time evolution of the same quantities, $B^2$ and $v^2$.
       The lower panel shows the time evolution of the standard deviation
      of the magnetic field (the dashed line)  and the passive scalar field  (the solid line) distributions.
                From Run 512-REF.
         }\label{fig:k20t}
    \end{minipage}
\end{figure}

\section{Discussions}

\subsection{Primordial or Astrophysical?}
 Our simulations show that the strength of a uniform seed magnetic field should be 
larger than $O(10^{-11})$G if $Re \sim 160$ in the ICM\footnote{
   In case $Re$ is smaller than $160$, the minimum strength of the seed magnetic 
   field will be larger than $O(10^{-11})$G.
   If $Re \sim 28$ as in Equation (\ref{eq:reicm}), we expect that the minimum strength
   will be at least $O(10^{-10})$G.}.
Note that observations of the Cosmic Microwave Background (CMB) place an upper limit 
of $O(10^{-9})$G on the strength of the uniform component of a primordial seed magnetic field
(Barrow, Ferreira \& Silk 1997; see also Widrow et al. 2012; Durrer \& Neronov 2013).
Therefore, it is unlikely that primordial magnetic fields are the direct
origin of present-day magnetic fields in the ICM.

If the high magnetic Prandtl number model is correct for fluids in the ICM, it is also unlikely that
a magnetic field  generated by the Biermann battery ($B\sim 10^{-20}$G) or
aperiodic turbulent fluctuations (Schlickheiser 2012; $B\sim 10^{-16}$G) are the
direct source of magnetic fields in the ICM. 
However, it is still possible that pre-galactic magnetic fields generated by the battery effect or the aperiodic turbulent fluctuations became
amplified in galaxies (see Kulsrud \& Zweibel 2008
and references therein; see also discussions in Beck \& Wielebinski 2013) or first stars 
(Sur et al. 2010; Schleicher et al. 2010; Schober et al. 2012b; Schleicher et al. 2013)
and then ejected into the intergalactic space later.
Such ejected magnetic fields can easily be the origin of cosmic magnetic fields 
even in high magnetic Prandtl number turbulence.
All in all, our simulations favor astrophysical origin of the ICM magnetic fields.

\subsection{Homogenization rate}
Cho \& Yoo (2012) showed that expansion of the magnetized region is fast in unit magnetic Prandtl number turbulence: 
homogenization of a doughnut-shaped seed magnetic field in unit magnetic Prandtl number turbulence
occurs within $\sim$3(L/v).
The numerical setup in Cho \& Yoo (2012) is almost identical to our current numerical setup for
the doughnut-shaped seed magnetic fields, 
excepts the magnetic Prandtl numbers (i.e. the values of the viscosity).
Our current simulations show that expansion of magnetized region is also fast in high Prandtl number turbulence: 
homogenization of a tube-like seed field occurs within $\sim 3(L/v)$ and
homogenization of a doughnut-like seed field occurs within $\sim 4(L/v)$.
In our current simulations, the large viscosity damps small-scale velocity.
Nevertheless, we obtain fast homogenization in our current simulations. Why is it so?
It seems that expansion of the magnetized region is governed by eddy motions at the outer scale of turbulence.
Inside an outer-scale eddy, magnetization of the whole eddy happens roughly within one large-scale eddy turnover time
if the size of the initially magnetized region is not much smaller than the viscous-cutoff scale.
After full magnetization of the outer-scale eddy, the magnetic field is transported over uncorrelated outer-scale eddies.
Because any outer-scale eddy which is partially magnetized can be fully magnetized within roughly
one large-scale  eddy turnover time, 
the rate of expansion of the magnetized region should be proportional to $v$ (Cho 2013).
Therefore, full magnetization of the system occurs within $\sim L_{sys}/v$, which we actually observe in our simulations.

Can we numerically confirm that the expansion speed of the magnetized region is indeed of order $v$
in high magnetic Prandtl number turbulence?
To measure the expansion rate on scales larger than the outer-scale eddies, 
we perform a numerical simulation with a small-scale driving (512-REF in Table 1).
Simulation setup for 512-REF is identical to that for 512-R1-DB$_{max}10^{-4}$ except the
the driving scale: the driving wave-numbers for 512-REF are between 15.4 and 26.
The hydrodynamic Reynolds number $Re$ for 512-REF is an order of magnitude smaller than that for 512-R1-DB$_{max}10^{-4}$ because
the driving scale for 512-REF is an order of magnitude smaller.
We plot the simulation results in Figures~\ref{fig:k20jpg} and \ref{fig:k20t}.
Figure~\ref{fig:k20jpg} shows the expansion of the magnetized region and
Figure~\ref{fig:k20t} shows the time evolution of $B^2$, $v^2$, and the standard deviations.
If we compare the rates of expansion for $0\leq t\leq 8.7$ and for $8.7\leq t\leq 17.4$
in Figure~\ref{fig:k20jpg},
the rates look similar, which implies that the expansion rate is linear in time.
Indeed the lower panel of Figure~\ref{fig:k20t} shows that the standard deviation of magnetic field distribution
(the dashed curve; see Equation~(\ref{eq:sig3d}) for definition) shows a linear growth after a certain time, before which
its behavior is very similar to that of a passive scalar field (the solid curve). 
This behavior is very similar to that of a localized seed magnetic field in unit magnetic Prandtl number turbulence
(see Cho 2013 for numerical methods and detailed discussions).

\subsection{Discussions on the localized seed magnetic fields}
If the initially magnetized region is smaller than the viscous-cutoff scale $l_d \sim Re^{-3/4}L$, the expansion speed
of the magnetized region can be slower than $\sim v$.
Let the size of the initially magnetized region be $D_s$.
If $D_s < l_d$, then it takes 
\begin{equation}
      \sim \frac{ l_d }{ v_d } \left( \frac{ l_d }{ D_s } \right)^2 \sim Re^{-2} 
                     \left( \frac{ L  }{ D_s }\right)^2 \left( \frac{ L }{ v } \right)
\end{equation}
for the magnetic field
to fill the `host' eddy at the viscous-cutoff scale.
After filling the viscous-cutoff-scale eddy, it would take $\sim L/v$ for the magnetic field
to fill the whole outer-scale eddy.
Therefore, in this case, magnetization of the outer-scale eddy takes
\begin{equation}
    \sim \left( 1+ Re^{-2} (L/D_s)^2 \right) \left( \frac{ L }{ v } \right).
\end{equation}
The expansion rate of the magnetized region on scales larger than the outer scale of turbulence will be
\begin{equation}
   \sim  v/\left( 1+ Re^{-2} (L/D_s)^2 \right) \label{eq:slowdown}
\end{equation}
For a cluster with $L\sim$400 kpc and $D_s \sim$30 kpc, we will have $D_s < l_s$ if $Re < (L/D_s)^{4/3} \sim 30$.
But, for this $Re$, slowdown of the expansion rate will be negligible (see Equation (\ref{eq:slowdown})).
Slowdown of the expansion rate will be important when $Re < (L/D_s) \sim$13 in the cluster. 
Note, however, that, if the magnetized materials in jets or stripped gases provide seeding, the initially 
magnetized region can be up to hundreds-of-kpc-long and, therefore, $Re$ can be smaller than this.

In the ISM of galaxies, strength of the magnetic fields ranges from $\sim \mu$G (diffuse ISM) to $\sim $mG (molecular clouds).
When these gases are expelled from galaxies, magnetic fields in them will be attenuated due to expansion of the media.
This makes the strength of the localized seed magnetic fields weaker than the galactic values.
If a linear size of the medium expands by a factor of $f_e$ during the ejection, magnetic field becomes weaker by a factor of 
$f_e^{-2}$.
The magnetic attenuation factor $f_e^{-2}$ is uncertain and may depend on the ejection mechanism.
If a galactic material is ejected via jets or explosions, the expansion factor can be very large.
For example, if a jet is launched from a sub-pc scale and ultimately expands to kpc-scales,
the expansion factor can be larger than $10^3$ (see related discussions in  Brandenburg \& Subramanian 2005; Widrow et al. 2012).
However, if a magnetized gas is stripped from a galaxy, the expansion factor can be smaller than $10^3$. 
Another factor that can affect the strength of the seed fields is existence of turbulence inside ejecta.
If turbulence exists inside the ejected media, turbulence dynamo can mitigate attenuation of the magnetic fields.

\subsection{Observational implications}
If we use the turbulence dynamo models, we can test whether or not the high magnetic Prandtl number turbulence
is a correct model for the ICM or the intergalactic medium.
Suppose that we can observe distribution of magnetic fields in filaments.
If distribution of magnetic fields in filaments is more or less homogeneous, it is likely that the 
high magnetic Prandtl number turbulence model is incorrect.
If we find magnetic fields only near the astrophysical sources, we may be able to conclude
that magnetic fields in filaments have astrophysical origins, rather than primordial.

\subsection{Effects of compression}
 The use of an incompressible code has the advantage that we can easily control the 
viscosity and the magnetic diffusivity.
However, it is not possible to study the effects of compression with the incompressible code.
In galaxy clusters, shocks and mergers can compress fluids and 
have a pronounced impact on the growth of magnetic field (Roettiger et al 1999; Dolag et al. 1999, 2002;
Iapichino \& Br{\"u}ggen 2012).
For example, a simulation by Dolag et al. (2002) shows that the mean magnetic field 
within a cluster of comoving radius 1 Mpc increases by a factor of $\sim 30$ from redshift
z=0.8 to z=0.
If this is a typical growth factor for compression amplification,
the minimum strength of a uniform magnetic field for $Re$=160 should be decreased by
the same factor: a new minimum strength will be a few times
$10^{-13}$G.
Note, however, that $Re$ in actual clusters is likely to 
be much smaller than $160$ (see Equation (\ref{eq:reicm})).
Therefore, if we take into account both the compression amplification and a realistic $Re$,
the minimum strength may not be much smaller than $O(10^{-11})$G.

It is likely that the central region of a galaxy cluster has formed at a much earlier stage. 
During these early stages, many of the physical properties, including the hydrodynamic
Reynolds number $Re$, could be different. 
It is therefore conceivable that some magnetization has occurred already at earlier stages,
which can alleviate the constraints found in this paper.
However, unless such central regions were substantially cooler and/or denser than present-day
central regions, the effect may not be significant.
Note that, even in present-day cool cores, $Re$ is expected to be
less than $\sim 10^2$ (Schekochihin \& Cowley 2006).

\section{Summary}
In this paper, we have considered turbulence dynamo models in
unit and high magnetic Prandtl number fluids.
We have found the following results:

\begin{enumerate}

\item In high magnetic Prandtl number turbulence (with a hydrodynamic Reynolds number, $Re$,
 less than $O(10^2)$),
      turbulence dynamo is not so efficient  
       that a primordial seed magnetic field weaker than a certain critical value cannot
       be a direct origin of magnetic fields in the ICM.
       For a cluster with driving scale of $\sim$400 kpc and turbulence velocity of $\sim$400 km/s,
       the critical strength is $\sim 10^{-11}$G for $Re=160$, which is very close to an
       upper limit of $O(10^{-9})$G placed by the CMB anisotropy observations.
       The critical strengths for seed magnetic fields ejected from galaxies or first stars are 
       about two orders of magnitude higher.
       But, since the strengths of magnetic fields in those astrophysical objects are likely to be larger than $\mu$G,
       it may not be so difficult
       for the seed magnetic fields ejected from the astrophysical bodies to have strengths larger than the critical values. 
        Therefore, our calculations favor astrophysical origin of magnetic fields in the ICM.
        
\item If the high magnetic Prandtl number model is correct for clusters and filaments,
        pre-galactic magnetic fields generated by a battery effect, amplified in galaxies or first stars, 
        and ejected from them later into
        intergalactic space would be plausible sources of magnetic fields in the large-scale
        structure of the universe.

\end{enumerate}

\acknowledgements
We thank the referee for useful suggestions and comments.
This  work is supported by the National R \& D Program through 
the National Research Foundation of Korea, 
funded by the Ministry of Education (NRF-2013R1A1A2064475).
This work is also supported by the Supercomputing Center/Korea Institute of Science and Technology Information with supercomputing resources including technical supports (KSC-2013-C1-034).

\begin{deluxetable}{lccccrrr}
\tabletypesize{\scriptsize}
\tablecaption{Simulations.}
\tablewidth{0pt}
\tablehead{
\colhead{Run} & \colhead{Resolution} & \colhead{Reynolds No.}
   & \colhead{$B_{max}$ at t=0\tablenotemark{a}} 
   & \colhead{$B_0$\tablenotemark{b}} &  \colhead{$Pr_m$\tablenotemark{c}}  & \colhead{Seed Field}
   & \colhead{ $L_{sys}/L$\tablenotemark{d} }
}
\startdata
512-R1-UB$_010^{-2}$        & $512^3$ & $\sim$160 & -  & $10^{-2}$ &  high & uniform & $\sim2.5$ \\ 
512-R1-UB$_010^{-4}$        & $512^3$ & $\sim$160 & -  & $10^{-4}$ &  high & uniform & $\sim2.5$ \\ 
512-R1-UB$_010^{-6}$        & $512^3$ & $\sim$160 & -  & $10^{-6}$ &  high & uniform & $\sim2.5$ \\ 
\hline
512-R2-UB$_010^{-4}$        & $512^3$ & $\sim$830 & -  & $10^{-4}$ &  high & uniform & $\sim2.5$ \\ 
512-R2-UB$_010^{-6}$        & $512^3$ & $\sim$830 & -  & $10^{-6}$ &  high & uniform & $\sim2.5$ \\ 
512-R2-UB$_010^{-8}$        & $512^3$ & $\sim$830 & -  & $10^{-8}$ &  high & uniform & $\sim2.5$ \\ 
\hline
512-R1-TB$_{max}10^{-2}$ & $512^3$ & $\sim$160 &  $10^{-2}$  & -  & high & tube-shaped & $\sim2.5$ \\ 
512-R1-TB$_{max}10^{-4}$ & $512^3$ & $\sim$160 &  $10^{-4}$  & -  & high  & tube-shaped & $\sim2.5$ \\ 
512-R1-DB$_{max}10^{-2}$ & $512^3$ & $\sim$160 & $10^{-2}$  & - & high & doughnut-shaped & $\sim2.5$ \\ 
512-R1-DB$_{max}10^{-4}$ & $512^3$ & $\sim$160 &  $10^{-4}$  & - &  high & doughnut-shaped & $\sim2.5$ \\
512-REF                  & $512^3$ & $\sim$10 &  $10^{-4}$  & - &  high & doughnut-shaped & $\sim20$ 
\enddata
\tablenotetext{a}{Maximum strength of the localized magnetic field at t=0.}
\tablenotetext{b}{Strength of the uniform magnetic field.}
\tablenotetext{c}{Magnetic Prandtl number ($\equiv \nu/\eta$).}
\tablenotetext{d}{$L_{sys}$ is the system size and $L$ is the driving scale.}
\label{table_1}
\end{deluxetable}

\appendix
\section{Turbulence dynamo model in unit magnetic Prandtl number fluids}
In this Appendix, we briefly review turbulence dynamo models in unit magnetic Prandtl number turbulence and
discuss their implications.
We assume that the fluid is incompressible and that
both the viscosity and the magnetic diffusivity are very small.
When the magnetic Prandtl number is unity, the velocity and the magnetic dissipation scales
coincide.

 If a seed magnetic field has a primordial origin,
the coherence length of the seed field can be very large and, therefore, it can be regarded as spatially uniform 
 on the scale of galaxy clusters.
 Even if the primordial seed magnetic field has a small coherence length (see, for example, Banerjee \& Jedamzik 2004; Wagstaff et al. 2014), turbulence dynamo for the seed field will be similar to that for a uniform seed field, 
 as long as the seed field is spatially homogeneous.
 On the other hand, 
 if a seed magnetic field was provided by an astrophysical object, it is likely that
the seed magnetic field was spatially localized at the time of injection.

\subsection{Amplification of a uniform seed field}
This topic has been studied extensively in literature.
In summary,  growth of a uniform weak seed magnetic field in unit magnetic Prandtl number turbulence
follows the following three stages (see Schl\"{u}ter \& Bierman 1950; Schekochihin \& Cowley 2007; Cho et al. 2009; see also
Cho \& Vishniac 2000;  Schober 2012a).

\begin{figure}
\center
\includegraphics[width=0.48\textwidth, bb=80 0 620 550]{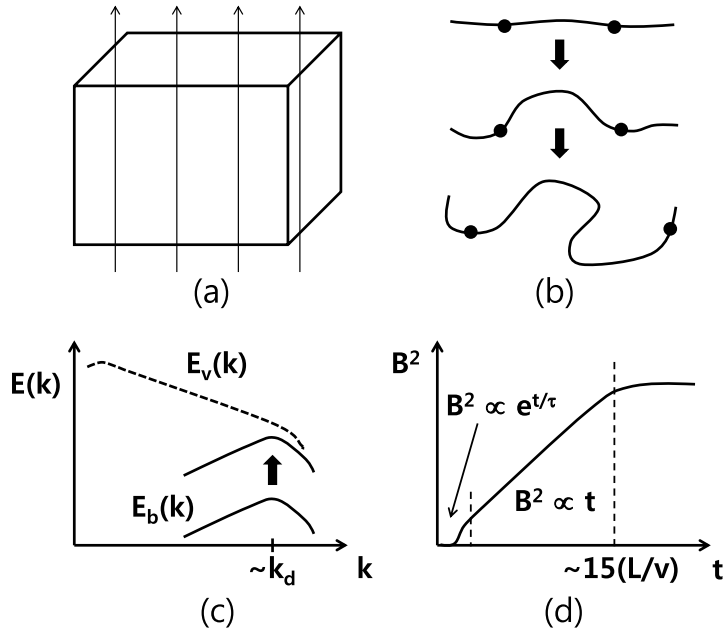}  
\caption{Turbulence dynamo for a uniform seed magnetic field. 
    We assume unit magnetic Prandtl number with a very large hydrodynamic Reynolds number.
  (a) If the seed magnetic field is a primordial one, it can be regarded as
  spatially uniform or homogeneous on comoving scales of $\sim$Mpc.
  (b) When we consider two points whose separation is $\sim l_d$,
       the distance between the two point along the field line doubles
       after each eddy turnover time, $\sim l_d/v_d$, where $v_d$ is velocity
       at the velocity dissipation scale $l_d$.
       Therefore, the distance grow exponentially, so does magnetic energy density.
   (c) Since the characteristic scale in (b) is $l_d$, the magnetic energy spectrum
        peaks near the dissipation wavenumber $k_d$. As time goes on, the magnetic
        spectrum goes up without changing the shape much.
   (d) In summary, the magnetic field grows in 3 steps: exponential growth, linear growth, and
         saturation.
         When the hydrodynamic Reynolds number ($Re$) approaches infinity, the duration of the exponential growth 
         stage becomes zero.
 }
\label{fig:1}
\end{figure}

(1) The stretching
of the magnetic field lines occurs most actively near the
velocity dissipation scale first.
 This is because eddy turnover time is shortest, hence the rate of field-line stretching is
 highest,
at the dissipation scale. Suppose that we introduce a weak uniform seed magnetic field in
a turbulent medium (see Figure \ref{fig:1}(a)).  Let us select two points on a magnetic field line whose separation
is of order $l_d$, where $l_d$ is the velocity dissipation scale (Figure \ref{fig:1}(b)).
After approximately one eddy turnover time, $\tau_d \sim l_d/v_d$, where $v_d$ is the velocity
at the dissipation scale, the distance between the two points along the field line will be doubled.
After approximately another eddy turnover time $\tau_d$, the distance between the two points along the field line
will be doubled again. Therefore, magnetic field lines are stretched exponentially and, as a result,
the magnetic energy density grows exponentially.
Note that the typical scale of the magnetic field-line variation, which is of order $l_d$,  does not change during the process. 
Therefore, the magnetic energy spectrum peaks near $k_d~(\propto 1/l_d)$ and, as time goes on, it goes upward
without changing its shape much (Figure \ref{fig:1}(c)).

(2) As the magnetic energy spectrum moves upward, it will finally touch up the kinetic
energy spectrum at the dissipation scale. Then what will happen? 
When the magnetic energy spectrum becomes comparable to the kinetic energy spectrum at the dissipation scale, 
magnetic back-reaction suppresses stretching of magnetic field lines at the dissipation scale.
Due to the suppression, the exponential growth
stage ends. Note, however, that stretching at scales  
larger than $l_d$ is still efficient because the kinetic energy spectrum is still higher than 
the 
magnetic energy spectrum at the scales. Therefore, now the scale slightly larger than the dissipation scale
becomes the most active scale for stretching. This way,  the subsequent stage
is characterized by a slower growth of magnetic energy and a
gradual shift of the peak of the magnetic energy spectrum to larger
scales (Cho \& Vishniac 2000). The growth rate of the magnetic energy density 
at this stage turns out to be linear (Schl\"{u}ter \& Biermann 1950; Schekochihin \& Cowley 2007). 

   (3) The amplification of magnetic field stops when
the magnetic energy density becomes comparable to the kinetic
energy density at the outer-scale of turbulence and a final, statistically steady, saturation stage
begins. During the saturation stage,  the peak of the magnetic energy spectrum occurs at a wavenumber
a few times larger than that of the kinetic energy spectrum (Cho \& Vishniac 2000).

In the vanishing viscosity (and diffusivity) limit, the duration of the exponential growth stage
will be very short. Nevertheless, since the eddy turnover time at the dissipation scale is arbitrarily small in
the limit, any weak seed field can grow exponentially, reach the equipartition strength at the
dissipation scale, and enter the linear growth stage within arbitrarily small amount of time.
Therefore, in the  vanishing viscosity limit, 
    1) the exponential growth stage shown in Figure \ref{fig:1}(d) becomes invisible, 
    2) $B^2(t)$ shows linear growth most of the time before saturation, 
and 3) regardless of the strength of the initial seed
field, $B^2(t)$ follows virtually the same curve of growth when $B^2$ is plotted in
a linear scale as in Figure \ref{fig:1}(d).
According to simulations (e.g. Cho et al. 2009; Cho \& Yoo 2012), the system reaches the saturation stage
in $\sim 15 (L/v)$.

\subsection{Amplification of a localized seed field}

\begin{figure}
\center
\includegraphics[width=0.40\textwidth, bb=130 250 520 540]{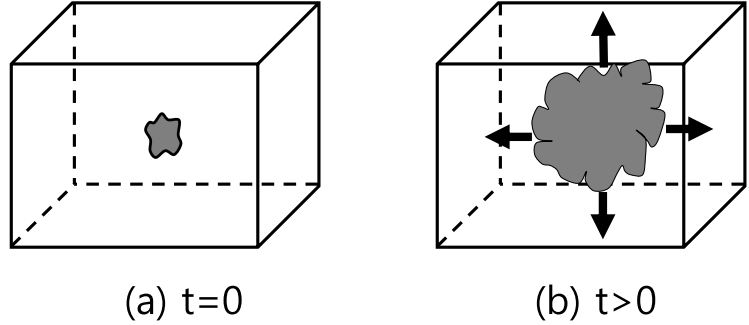}  
\caption{A localized seed magnetic field in a turbulent medium.
 (a) If the seed magnetic field is ejected from an astrophysical body, it can be
  spatially localized at the time of injection.
  (b) As time goes on, the magnetized region expands.
     The speed of expansion is $\sim v$, where $v$ is the turbulence velocity.
 }
\label{fig:2}
\end{figure}

Amplification of a localized seed magnetic field was studied in detail by Cho \& Yoo (2012) and Cho (2013).
If a seed magnetic field is localized in space, turbulent motions make the seed magnetic field
transported out from the region of initial injection (Figure~\ref{fig:2}).
Cho (2013) found that the rate at which the magnetized region expands is $\sim v$.
Therefore, the {\it homogenization timescale} (or the magnetization timescale),
the time required for the magnetic field to fill the whole system, is given by
\begin{equation}
   t_{mag} \sim L_{sys}/v.
\end{equation}
If the driving scale is comparable to the size of the system, $L_{sys}$, then
homogenization happens fast, 
i.e.~within $L_{sys}/L$ times the large-eddy turnover time $L/v$,
and the subsequent evolution should be similar to
that of a uniform/homogeneous seed magnetic field case (Cho \& Yoo 2012)\footnote{
    This result is consistent with earlier cosmological simulations.
   Dolag et al. (1999, 2002), for example, found that information on the
   initial magnetic fields  (homogeneous or chaotic) is completely wiped out 
   during the cluster formation and simulations
    yield similar results. 
}.
The whole system reaches the saturation stage within $\sim 15 L_{sys}/v$.
On the other hand, if the driving scale is less than $\sim L_{sys}/15$,
 i.e.~if $L_{sys}/L >15$,
the magnetic field near the initial injection cite reaches 
the equipartition strength $B_{eq}$   
before
the magnetic field fills the whole system. In this case, it takes more than $\sim L_{sys}/v =(L_{sys}/L) (L/v)$
for the whole system to reach the saturation stage.

As in uniform/homogeneous seed field cases, as
long as the strength of the seed magnetic field is sufficiently weaker than the equipartition strength,
its absolute value does not matter much in the  vanishing viscosity limit.
First, because weak magnetic fields are passively transported and stretched by
turbulent motions, the homogenization timescale $t_{mag}$, which 
is mainly determined by large-scale turbulent motions,
is similar for all weak seed magnetic fields.
Second, as in the uniform/homogeneous seed field cases,
     the strength of the weak seed fields does not affect 
     the saturation time much.  
Therefore, the turbulence dynamo process should be similar for sufficiently
weak localized seed magnetic fields.

\subsection{Observational Implications}
Since turbulence dynamo is so efficient 
in the vanishing viscosity limit that
it is difficult to find a constraint on the strength of the initial seed magnetic field.
This is true for both the primordial and the astrophysical seed magnetic fields.

However, it might be possible to tell the origin of the cosmic magnetism by observing the distribution of magnetic field in some systems.
The key quantity is
the homogenization timescale, $t_{mag}$.
Suppose that a localized seed magnetic field is injected into a system.
If $t_{mag}$ is short compared with the age of the universe,
the system becomes homogenized quickly and the subsequent evolution
will be very similar to that of a uniform seed magnetic field.
Therefore, in this case,
it will be very difficult to distinguish the astrophysical and the primordial seed magnetic fields.
On the other hand, if $t_{mag}$ is longer than the Hubble time,
the system has not been yet completely magnetized. Therefore, in this case, the magnetic field in the system will
be very inhomogeneous.

Consider a cluster of galaxies with $L_{sys} \sim$1 Mpc.
In this case, if $v > 75$ km/s, which is very likely, the homogenization timescale is shorter than
the Hubble time. 
Therefore, it will be very difficult to distinguish the primordial and the astrophysical origins
of magnetic fields in clusters.
On the other hand, if we consider a filament of width $\sim$4 Mpc, 
the homogenization timescale
is longer than the Hubble time, if $v<$300 km/s.
Therefore, it is possible that filaments have not been fully homogenized.
If this is the case, the distribution of magnetic fields in filaments is more or less inhomogeneous.


\begin{thebibliography}{99}   

\bibitem[Arieli et al.(2011)]{2011ApJ...738...15A} Arieli, Y., Rephaeli, 
Y., \& Norman, M.~L.\ 2011, \apj, 738, 15 



\bibitem[Banerjee 
\& Jedamzik(2004)]{2004PhRvD..70l3003B} Banerjee, R., \& Jedamzik, K.\ 2004, \prd, 70, 123003 




\bibitem[Name (1997)]{Bar97} Barrow, J., Ferreira, P., \& Silk, J.  1997, Phys. Rev. Lett., 78, 3616

\bibitem[Name (2013)]{Bat50} Batchelor, G.\ 1950, Proc. R. Soc. London A, 201, 405 

\bibitem[Beck 
\& Wielebinski(2013)]{2013pss5.book..641B} Beck, R., \& Wielebinski, R.\ 2013, Planets, Stars and Stellar Systems.~Volume 5: Galactic Structure and Stellar Populations, 641


\bibitem[Beresnyak(2012)]{2012PhyS...86e8201B} Beresnyak, A.\ 2012, 
\physscr, 86, 058201 

\bibitem[Biermann(1950)]{1950ZNatA...5...65B} Biermann, L.\ 1950, 
Zeitschrift Naturforschung Teil A, 5, 65 

\bibitem[Bovino et al.(2013)]{2013NJPh...15a3055B} Bovino, S., Schleicher, 
D.~R.~G., \& Schober, J.\ 2013, New Journal of Physics, 15, 013055 


\bibitem[Brandenburg et al.(2012)]{2012SSRv..169..123B} Brandenburg, A., 
Sokoloff, D., \& Subramanian, K.\ 2012, \ssr, 169, 123 

\bibitem[Name (2013)]{BraS05} Brandenburg, A., \& Subramanian, K.\ 2005,   Phys. Reports, 417, 1

\bibitem[Brunetti 
\& Lazarian(2007)]{2007MNRAS.378..245B} Brunetti, G., \& Lazarian, A.\ 2007, \mnras, 378, 245 


\bibitem[Carilli 
\& Taylor(2002)]{2002ARA&A..40..319C} Carilli, C.~L., \& Taylor, G.~B.\ 2002, \araa, 40, 319 

\bibitem[Childress(1995)]{ChiG95} Childress, S., \& Gilbert, A.\ 1995, Stretch, Twist, Fold: The Fast Dynamo (Berlin: Springer)

\bibitem[Cho(2013)]{2013PhRvD..87d3008C} Cho, J.\ 2013, \prd, 87, 043008 

\bibitem[Name (2013)]{ChoV00} Cho, J., \& Vishniac, E.~T.\ 2000, \apj, 538, 217

\bibitem[Cho et al.(2009)]{ChoVB09} Cho, J., Vishniac, E.~T., 
Beresnyak, A., Lazarian, A., \& Ryu, D.\ 2009, \apj, 693, 1449 

\bibitem[Name (2013)]{ChoY12}Cho, J., \& Yoo, H.\ 2012,  \apj, 759, 91

\bibitem[Daly 
\& Loeb(1990)]{1990ApJ...364..451D} Daly, R.~A., \& Loeb, A.\ 1990, \apj, 364, 451 

\bibitem[Donnert et al.(2009)]{2009MNRAS.392.1008D} Donnert, J., Dolag, K., 
Lesch, H., M\''{u}ller, E.\ 2009, \mnras, 392, 1008 

\bibitem[Dolag et 
al.(1999)]{1999A&A...348..351D} Dolag, K., Bartelmann, M., \& Lesch, H.\ 1999, \aap, 348, 351 

\bibitem[Dolag et 
al.(2002)]{2002A&A...387..383D} Dolag, K., Bartelmann, M., \& Lesch, H.\ 2002, \aap, 387, 383 


\bibitem[Durrer \& Neronov(2013)]{2013A&ARv..21...62D} 
        Durrer, R., \& Neronov, A.\ 2013, \aapr, 21, 62 
        
\bibitem[Ferrari et al.(2008)]{2008SSRv..134...93F} Ferrari, C., Govoni, 
F., Schindler, S., Bykov, A.~M., \& Rephaeli, Y.\ 2008, \ssr, 134, 93 







\bibitem[Haugen et al.(2004)]{2004PhRvE..70a6308H} Haugen, N.~E., 
Brandenburg, A., \& Dobler, W.\ 2004, \pre, 70, 016308 

\bibitem[Hoyle(1969)]{1969Natur.223..936H} Hoyle, F.\ 1969, \nat, 223, 936 

\bibitem[Iapichino 
\& Br{\"u}ggen(2012)]{2012MNRAS.423.2781I} Iapichino, L., \& Br{\"u}ggen, M.\ 2012, \mnras, 423, 2781 

\bibitem[Govoni 
\& Feretti(2004)]{2004IJMPD..13.1549G} Govoni, F., \& Feretti, L.\ 2004, International Journal of Modern Physics D, 13, 1549 

\bibitem[Name (2013)]{Kaz68} Kazantsev, A.~P.\ 1968, Soviet Phys.-JETP Lett., 26, 1031 

\bibitem[Name (2013)]{Kro94} Kronberg, P.~P.\ 1994,  
              Reports on Progress in Physics, 57, 325
             
\bibitem[Kronberg et al.(2001)]{2001ApJ...560..178K} Kronberg, P.~P., 
Dufton, Q.~W., Li, H., \& Colgate, S.~A.\ 2001, \apj, 560, 178 

\bibitem[Kulsrud 
\& Anderson(1992)]{KulA92} Kulsrud, R.~M., \& Anderson, S.~W.\ 1992, \apj, 396, 606 

\bibitem[Kulsrud et al.(1997)]{Kul97} Kulsrud, R.~M., Cen, 
R., Ostriker, J.~P., \& Ryu, D.\ 1997, \apj, 480, 481 

\bibitem[Name (2013)]{KulZ08} Kulsrud, R.~M., \& Zweibel, E.~G.\ 2008
           Reports on Progress in Physics, 71, 046901
           
\bibitem[Meneguzzi et al.(1981)]{1981PhRvL..47.1060M} Meneguzzi, M., 
Frisch, U., \& Pouquet, A.\ 1981, Physical Review Letters, 47, 1060 

\bibitem[Miniati(2014)]{2014ApJ...782...21M} Miniati, F.\ 2014, \apj, 782, 
21 

\bibitem[Name (2013)]{MolRS85} Molchanov, S., Ruzmaikin, A., \& Sokolov, D.\ 1985,  Sov. Phys. Usp., 28, 307


\bibitem[Pouquet et al.(1976)]{1976JFM....77..321P} Pouquet, A., Frisch, 
U., \& Leorat, J.\ 1976, Journal of Fluid Mechanics, 77, 321 

\bibitem[Pudritz 
\& Silk(1989)]{1989ApJ...342..650P} Pudritz, R.~E., \& Silk, J.\ 1989, \apj, 342, 650 

 \bibitem[Rephaeli(1988)]{1988ComAp..12..265R} Rephaeli, Y.\ 1988, Comments 
on Astrophysics, 12, 265 


\bibitem[Rees(1987)]{Rees87}  Rees, M.~J.\ 1987
         Royal Astronomical Society, Quarterly Journal, 28, 197 
         
\bibitem[Name (2013)]{Rey05} Reynolds, C., McKernan, B., Fabian, A., Stone, J., \& Vernaleo, J.\ 2005, \mnras, 357, 242

\bibitem[Robinson et al.(2004)]{2004ApJ...601..621R} Robinson, K., Dursi, 
L.~J., Ricker, P.~M., et al.\ 2004, \apj, 601, 621 

\bibitem[Roettiger et al.(1999)]{1999ApJ...518..594R} Roettiger, K., Stone, 
J.~M., \& Burns, J.~O.\ 1999, \apj, 518, 594 

\bibitem[Name (2013)]{Rus04} Ruszkowski, M., Br\"uggen, M., \& Begelman, M.\ 2004,  \apj, 611, 158

\bibitem[Name (2013)]{RuzSS89}  Ruzmaikin, A., Sokoloff, D., \& Shukurov,  A.\ 1989, \mnras, 241, 1

\bibitem[Name (2013)]{Ryu08} Ryu, D., Kang, H., Cho, J., \& Das, S.\ 2008, Science, 320, 909

\bibitem[Ryu et al.(2012)]{2012SSRv..166....1R} Ryu, D., Schleicher, 
D.~R.~G., Treumann, R.~A., Tsagas, C.~G., 
\& Widrow, L.~M.\ 2012, \ssr, 166, 1 

\bibitem[Schekochihin 
\& Cowley(2006)]{2006PhPl...13e6501S} Schekochihin, A.~A., \& Cowley, S.~C.\ 2006, Physics of Plasmas, 13, 056501 

\bibitem[Name (2013)]{SchC07} Schekochihin, A., \& Cowley, S.\ 2007, in 
       {\it Magnetohydrodynamics - Historical evolution and trends},
        eds. by  Molokov, S.,  Moreau, R., \&  Moffatt, H.
       (Berlin; Springer), p.~85 (astro-ph/0507686)

\bibitem[Name (2013)]{Sch04} Schekochihin, A.~A., Cowley, S.~C., Taylor, S.~F., Maron, J.~L., \& McWilliams, J.~C.\ 2004, 
                 \apj, 612, 276
                
\bibitem[Schleicher et 
al.(2010)]{2010A&A...522A.115S} Schleicher, D.~R.~G., Banerjee, R., Sur, S., et al.\ 2010, \aap, 522, A115 

\bibitem[Schleicher et al.(2013)]{2013AN....334..531S} Schleicher, 
D.~R.~G., Latif, M., Schober, J., et al.\ 2013, Astronomische Nachrichten, 
334, 531 


\bibitem[Schlickeiser(2012)]{2012PhRvL.109z1101S} Schlickeiser, R.\ 2012, 
Physical Review Letters, 109, 261101 

\bibitem[Schl{\"u}ter 
\& Biermann(1950)]{1950ZNatA...5..237S} Schl{\"u}ter, A., \& Biermann, I.\ 1950, Zeitschrift Naturforschung Teil A, 5, 237 

\bibitem[Schober et al.(2012)]{2012PhRvE..85b6303S} Schober, J., 
Schleicher, D., Federrath, C., Klessen, R., 
\& Banerjee, R.\ 2012a, \pre, 85, 026303   

\bibitem[Schober et al.(2012)]{2012ApJ...754...99S} Schober, J., 
Schleicher, D., Federrath, C., et al.\ 2012b, \apj, 754, 99    






\bibitem[Spitzer(1962)]{1962pfig.book.....S} Spitzer, L.\ 1962, Physics of 
Fully Ionized Gases (New York: Wiley)

\bibitem[Name (2013)]{SubSH06}  Subramanian, K., Shukurov, A., \& Haugen,  N.\ 2006, \mnras, 366, 1437

\bibitem[Sur et al.(2010)]{2010ApJ...721L.134S} Sur, S., Schleicher, 
D.~R.~G., Banerjee, R., Federrath, C., 
\& Klessen, R.~S.\ 2010, \apjl, 721, L134

\bibitem[Vainshtein 
\& Ruzmaikin(1972)]{1972SvA....15..714V} Vainshtein, S.~I., \& Ruzmaikin, A.~A.\ 1972, \sovast, 15, 714 

\bibitem[Vazza et 
al.(2011)]{2011A&A...529A..17V} Vazza, F., Brunetti, G., Gheller, C., Brunino, R., \& Br{\"u}ggen, M.\ 2011, \aap, 529, A17 

\bibitem[Vogt 
\& En{\ss}lin(2005)]{2005A&A...434...67V} Vogt, C., \& En{\ss}lin, T.~A.\ 2005, \aap, 434, 67 

\bibitem[Wagstaff et al.(2014)]{2014PhRvD..89j3001W} Wagstaff, J.~M., 
Banerjee, R., Schleicher, D., \& Sigl, G.\ 2014, \prd, 89, 103001 

\bibitem[Widrow(2002)]{2002RvMP...74..775W} Widrow, L.~M.\ 2002, Reviews of 
Modern Physics, 74, 775 

\bibitem[Widrow et al.(2012)]{2012SSRv..166...37W} Widrow, L.~M., Ryu, D., 
Schleicher, D.~R.~G., et al.\ 2012, \ssr, 166, 37 

\bibitem[Xu et al.(2010)]{2010ApJ...725.2152X} Xu, H., Li, H., Collins, 
D.~C., Li, S., \& Norman, M.~L.\ 2010, \apj, 725, 2152 

\bibitem[Yoo 
\& Cho(2014)]{2014ApJ...780...99Y} Yoo, H., \& Cho, J.\ 2014, \apj, 780, 99 


\bibitem[Zel'dovich(1984)]{Zel84}
Zel’dovich, Ya. B., Ruzmaikin, A. A., Molchanov, S. A., \& Sokoloff, D. D.\
1984, J.\ Fluid Mech., 144, 1

\bibitem[Zweibel 
\& Heiles(1997)]{1997Natur.385..131Z} Zweibel, E.~G., \& Heiles, C.\ 1997, \nat, 385, 131 


\end{thebibliography}
\end{document}